\documentclass[12pt]{article}    
\usepackage{hyperref}
\usepackage{graphics}
\usepackage{epsfig}
\usepackage{epstopdf}
\usepackage{natbib}   
\usepackage{graphicx} 
\usepackage{amssymb}  
\usepackage{amsmath}
\usepackage{multirow} 
\usepackage{enumitem} 
\usepackage{color}     
\usepackage{indentfirst}  
\usepackage{booktabs}
\usepackage{caption} 
\usepackage{appendix}
\renewcommand\appendix{\section*{Appendix}}
\usepackage[gen]{eurosym}
\usepackage[scientific-notation=true]{siunitx}
\usepackage{dsfont}
\usepackage{bm}
\usepackage {soul}
\usepackage{dcolumn}
\newcolumntype{d}[1]{D{.}{.}{#1}} 
\begin{document}

\begin{center}
\section*{Modelling Global Fossil CO$_2$ Emissions with a Lognormal Distribution: A Climate Policy Tool}

\vskip 0.1in {\sc \bf Faustino Prieto$^a$, Catalina B. Garc\'ia-Garc\'ia$^b$\footnote{Corresponding author. E-mail address: cbgarcia@ugr.es (C.B. Garc\'ia-Garc\'ia).}, Rom\'an Salmer\'on G\'omez$^b$
\vskip 0.1in

{\small\it 
$^a$Quantitative Methods for Economics and Business, University of Cantabria, Avenida de los Castros s/n, 39005 Santander, Spain.\\
$^b$Quantitative Methods for Economics and Business, University of Granada, 
Poligono la Cartuja s/n, 18071 Granada, Spain.
}\\
}

\end{center}

\begin{abstract}\noindent
Carbon dioxide (CO$_2$) emissions have emerged as a critical issue with profound impacts on the environment, human health, and the global economy. The steady increase in atmospheric CO$_2$ levels, largely due to human activities such as burning fossil fuels and deforestation, has become a major contributor to climate change and its associated catastrophic effects. To tackle this pressing challenge, a coordinated global effort is needed, which necessitates a deep understanding of emissions patterns and trends. In this paper, we explore the use of statistical modelling, specifically the lognormal distribution, as a framework for comprehending and predicting CO$_2$ emissions. We build on prior research that suggests a complex distribution of emissions and seek to test the hypothesis that a simpler distribution can still offer meaningful insights for policy-makers. We utilize data from three comprehensive databases and analyse six candidate distributions (exponential, Fisk, gamma, lognormal, Lomax, Weibull) to identify a suitable model for global fossil CO$_2$ emissions. Our findings highlight the adequacy of the lognormal distribution in characterizing emissions across all countries and years studied. Furthermore, to provide additional support for this distribution,
we provide statistical evidence supporting the applicability of Gibrat's law to those CO$_2$ emissions. Finally, we employ the lognormal model to predict emission parameters for the coming years and propose two policies for reducing total fossil CO$_2$ emissions. Our research aims to provide policy-makers with accurate and detailed information to support effective climate change mitigation strategies.
\end{abstract}
\vskip 0.2in

\noindent {\bf Key Words}: Lognormal distribution; Climate change mitigation;
Global fossil CO$_2$ emissions; Gibrat's law

\section{Introduction}

Carbon dioxide (CO$_2$) emissions are a critical issue that has far-reaching impacts on the environment, human health, and the global economy. The concentration of CO$_2$ in the atmosphere has increased significantly in recent decades, primarily due to human activities such as burning fossil fuels, deforestation, and industrial processes. This increase in CO$_2$ emissions is a significant contributor to climate change, which has the potential to cause catastrophic effects such as rising sea levels, extreme weather events, and food and water shortages. Because CO$_2$ emissions are directly linked to economic activity, their reduction requires a coordinated global effort, including the adoption of cleaner technologies and practices, investment in renewable energy, and policy changes to incentivize low-carbon economic growth. In line with these objectives, the European Climate Law, which was formally adopted in June 2021, enshrines the ambitious commitment of the European Union (EU) to achieve climate neutrality by 2050. To accomplish this goal, the EU aims to implement a comprehensive set of policies and measures to reduce greenhouse gas emissions, with a particular focus on CO$_2$ emissions from various sources.

One of the ways to better understand and predict emissions is through modelling as a framework for analysing data, making predictions and identifying trends and patterns that are not readily apparent through observation alone. \cite{hadley2003assessing} presented an interesting review of the different attempts to find the underlying distribution of different pollutant concentration levels. They concluded that a 2-parameter lognormal distribution can be a very good description of annual mean daily sulphur dioxide concentrations at 10 monitoring sites in the United Kingdom over time periods of up to 40 years for a wide range of ambient levels, time periods and monitoring site types. \cite{anthoff2023testing} analysed the right tail of the distribution of the published social cost of carbon estimates and tested the Dismal Theorem. They confirmed the importance of considering the distribution of key variables to help develop tools for policy-making. In relation to CO$_2$ emissions, \cite{akhundjanov2017size} examined the size distribution of national carbon dioxide emissions on a sample of 210 countries and territories for the period 2000-2010. They concluded that the composite of Pareto and lognormal distributions fit remarkably well to the entire cross-sectional distribution of CO$_2$ emissions over different time periods, after comparing lognormal, double Pareto-lognormal, lognormal-upper tail Pareto, and Pareto tail-lognormal distributions. The parametric analysis reveals that the upper tail of CO$_2$ emissions is characterized by Zipf's law. The presence of a power law in the higher range suggests that a significant amount of CO$_2$ emissions originate from larger countries.

Recently, \cite{pena2022log} studied the parametric distribution of log-growth rates of CO$_2$ and CO$_2$ per capita emissions for 207 countries and territories, taking data from 1994 to 2010. This study seems to be the first attempt to define the probability density functions of those variables by comparing the normal, the asymmetric double Laplace normal, the exponential tails normal and a mixture of two or three normal distributions.

Against that background, the first goal of this work is to analyse the hypothesis that the fossil CO$_2$ emissions data, at the country level, can be described by a 2-parameter statistical model for the whole range of the distribution (all world countries). We argue that modelling with a simple distribution can be particularly useful in understanding CO$_2$ emissions. Indeed, the second goal of this paper is to provide a climate policy tool at the country level to convert a worldwide emission goal into national reduction targets. Thus, by providing policy-makers with more accurate and detailed information, these models can help inform decisions and policies that have the potential to reduce emissions and mitigate the effects of climate change.

The structure of the paper is as follows: Section \ref{methodology} begins by presenting the three databases that will be used to model the annual fossil CO$_2$ emissions for each country in the world. It continues by describing the methodology applied to obtain the empirical application in relation to the six analysed distributions (exponential, Fisk, gamma, lognormal, Lomax, Weibull) and the proposed criteria for the selection of one of them. A subsection of the methodology is focused on the analysis of Gibrat's law and the methodologies for its verification. On the theoretical basis established above, Section \ref{tool} presents two different policies for reducing total fossil CO$_2$ emissions. Empirical results are shown in Section \ref{results}. First, the whole range of the distribution (all world countries) is modelled using the selected distributions, with the lognormal distribution selected as a suitable model in all three datasets considered and in all years analysed. To provide additional support for this distribution, four methodologies are applied to analyse the verification of Gibrat's law, concluding with strong statistical evidence. Finally, the applicability of adjusting CO$_2$ emissions with a simple distribution is shown. First, we provide predictions for the parameters of the lognormal distribution for 2025, 2030 and 2035 for the different databases. Second, by describing the two policies (in terms of scale and inequality) for reducing total fossil CO$_2$ emissions proposed in the methodology, Section \ref{conclusion} summarizes the main findings of this paper.

\section{Material and methods}\label{methodology}

\subsection{Data}\label{data}

In this paper, we considered fossil carbon dioxide (CO$_2$) emissions data at the country level for all the ranges of the distribution (of all world countries) on a yearly basis and expressed them in megatonnes of CO$_2$ per year ($\text{MtCO}_2\text{/year)}$.

We used the following free publicly accessible databases that provide estimates of annual fossil CO$_2$ emissions for each country in the world:
\begin{enumerate}[leftmargin=5 mm,nolistsep]
\item EDGAR v7.0 database (Emissions Database for Global Atmospheric Research). Developed by the European Commission - Joint Research Centre (EC-JRC) jointly with other organizations \citep{Crippa2022a,Crippa2022b}.\\
File: \textit{EDGARv7.0\_FT2021\_fossil\_CO2\_booklet2022.xlsx}
(fossil\_CO2\_totals \_by\_country sheet),
for the period 1970-2021, released in September 2022.
\item GCB 2022 database (Global Carbon Budget). Produced by an international cooperation of more than a hundred scientists from 80 organizations and 18 countries \citep{Friedlingstein2022a,Friedlingstein2022b}. \\
File: \textit{National\_Fossil\_Carbon\_Emissions\_2022v0.1.xlsx} (Territorial Emissions sheet), for the period 1850-2021, published in November 2022.
\item CDIAC-FF 2022 database (Carbon Dioxide Information and Analysis Center). Maintained currently by Appalachian State University and formerly housed at Oak Ridge National Laboratory (ORNL) \citep{CDIACFF2022,Gilfillan2021}. \\
File: \textit{nation.1751\_2019.xlsx}, for the period 1751-2019, last access in November 2022.
\end{enumerate}

It can be noted that those three datasets show differences in the list of countries considered, in the period of years covered, in the methodology for estimating the variable of interest, and in their measurement units.
For this study, we selected all the years since 1970 (52 years in the EDGAR and GCB datasets and 50 years in the CDIAC-FF dataset). We
did not consider International Aviation and International Shipping data in the EDGAR dataset. We expressed all the data in megatonnes of CO$_2$ per year ($\text{MtCO}_2\text{/year}$) using the conversion factor for CO$_2$ emissions from carbon (1 kg C = 3.664 kg $\text{CO}_2$-eq., \citep{Shin2017,Pielke2009} in the cases of the GCB and CDIAC-FF datasets.

For reference, Table \ref{tab1} shows a comparison of the main empirical characteristics of the variable of
interest, corresponding to three selected years (1970, 2000 and 2019) from the three datasets considered. In particular, it shows the number of countries analysed (N), the fossil $\text{CO}_2$ emission estimates of the world's largest emitter, the minimum, the mean and standard deviation (in $\text{MtCO}_2$), and finally, the skewness and kurtosis of those countries' fossil CO$_2$ emissions.

For the year 1970, EDGAR has 208 countries, GCB has 199 countries, and CDIAC-FF has 184 countries. In the subsequent years 2000 and 2019, the number of countries varied slightly but was still in the range of 200.

Looking at the statistics for the same year, we observe that the three databases (EDGAR, GCB, and CDIAC-FF) generally report similar patterns in terms of maximum, minimum, mean, and standard deviation. However, there are slight differences in skewness and kurtosis values, indicating some variations in the shape of the distributions. Notably, there was a significant increase in the maximum emission values from 1970 to 2019, indicating a rise in CO$_2$ emissions over the years.

\begin{table}[htp]\footnotesize
\renewcommand{\tablename}{\footnotesize{Table}}
\caption{\footnotesize\label{tab1}Some relevant information about the datasets used. Years: 1970, 2000, 2019.}
\setlength{\tabcolsep}{2 mm}
\begin{tabular}{@{}l c c c c c c c @{}}
\toprule
Database&N& Maximum& Minimum&Mean&
Std. Dev.&Skewness& Kurtosis\\
&& ($\text{MtCO}_2$)& ($\text{MtCO}_2$)&($\text{MtCO}_2$)&
($\text{MtCO}_2$)&&\\
\midrule
Year: 1970&&&&&&&\\
EDGAR &208 & 4693.3 & 0.0008 & 75.0 & 365.1 &10.3& 121.8\\
GCB      &199 & 4339.7 & 0.0037 & 72.7 & 345.2 &10.0& 116.1\\
CDIAC-FF&184&4325.3& 0.0037 & 78.3 & 378.1 & 9.0&   90.9\\
\midrule
Year: 2000&&&&&&&\\
EDGAR &208 & 6004.4 & 0.0017 & 120.1 & 518.7 &8.8& 88.1\\
GCB      &217 & 6016.4 & 0.0073 & 113.7 & 504.7 &9.1& 94.7\\
CDIAC-FF&217&5685.9& 0.0037 & 110.3 & 483.4 & 8.9& 90.8\\
\midrule
Year: 2019&&&&&&&\\
EDGAR &208 & 11771.1 & 0.0020 & 176.4 & 915.7 &10.6& 124.6\\
GCB      &219& 10741.0 & 0.0073 &  163.6 & 838.5 &10.3& 119.3\\
CDIAC-FF&222&10504.1& 0.0073 & 157.7 &  813.4 &10.4& 121.1\\
\bottomrule
\end{tabular}
\end{table}

\subsection{Modelling the whole range of the distribution (all world countries)}

As mentioned in the introduction, for this analysis, we considered the hypothesis that the fossil CO$_2$ emissions data, at the country level, can be described by a 2-parameter statistical model in the whole range of the distribution (all world countries). That means, in the event this hypothesis could not be rejected that we would have a simple parametric model that would be very useful in this context.

First, we selected the following six well-known models: Exponential (denoted as EXP), Fisk (log-logistic, FSK), Gamma (GAM), Lognormal (LOG), Lomax (Pareto type II with zero location parameter, PA2) and Weibull (WEI) distributions.
Those models were selected among the so-called statistical size distributions \citep{Kleiber2003}, according to the continuous and strictly positive nature of the variable of interest (the ``size" of the annual fossil CO$_2$ emissions for each country in the world). For that, we took into account the positive values of the skewness shown in  Table \ref{tab1}, that ruled out symmetrical models (as the normal distribution, among others) and the parsimony principle for a statistical model. Table \ref{tabA1} shows the cumulative distribution function $F(x)$, the probability density function 
$f(x)$, the support and the parameters (with their restrictions) of those six models \citep{Johnson1994,Kleiber2003,Teulings2023}.

Then, we fitted those models to the data expressed in megatonnes of CO$_2$ per year (MtCO$_2$/year) in the whole range by using the maximum likelihood method \citep{Fisher1922}, with the log-likelihood function given by
\begin{equation}\label{loglik}
\log[L(\bm{\theta}; \mathbf{x})]=\displaystyle\sum_{i=1}^N \log f(x_i;\bm{\theta}),
\end{equation}
where $x_i,\;i=1,\dots,N$ is a sample of size $N$,
$f(x)$ and $\bm{\theta}$ are the probability density function and the vector of parameters of the model considered, respectively. Maximum likelihood estimation
$\bm{\hat\theta}$ is the one that maximizes that function (\ref{loglik}). For that, we used the \verb|R| package \verb|fitdistrplus| (with the function \verb|fitdist|), complemented with the \verb|R| package \verb|actuar| \citep{Delignette2015,Dutang2008}.

After that, we compared those six models by using the Akaike information criterion, $AIC=-2\log[L(\bm{\hat\theta}; \mathbf{x})]+2K$, where $\log[L(\bm{\hat\theta}; \mathbf{x})]$ is the log-likelihood of the model evaluated at the maximum likelihood estimates, $K$ is the number of parameters of the model, and a better fit is indicated by a lower $AIC$ value. For that comparison, instead of using those $AIC$ values directly, we obtained the minimum value of $AIC$ ($AIC_{min}$) for each year and dataset. We calculated the value of $\Delta=AIC-AIC_{min}$ for each model, and we classified those models into three groups: models with the best fit ($\Delta\leq 2$), models with relatively little support ($2<\Delta\leq 20$), and models with no empirical support ($\Delta>20$) \citep{Burnham2011,Prieto2022}.

We found that the lognormal distribution was the best model from among the distributions considered in the study in all three datasets considered and in all years analysed. Then, we tested the null hypothesis:
$H_0$: {\it the fossil} CO$_2$ {\it emissions data, at country level,
follow a 2-parameter lognormal model in the whole range of the distribution (all world countries)} by using seven different analytical tests and considering significance levels of 0.05 and 0.01.

Taking into account that if a random variable $X$ follows a lognormal distribution then its logarithm $\log(X)$ follows a normal distribution, we considered the following tests for normality: the Shapiro--Wilk (SW), Shapiro--Francia (SF), Lilliefors (LL), Cramer--von Mises (CVM), Anderson--Darling (AD), D'Agostino--Pearson Omnibus Test (DP) and Jarque--Bera (JB) tests \citep{Seier2002,Razali2011,Yap2011}. Notably, the two first tests are regression and correlation tests, the three following tests are based on the empirical distribution function, and the last two tests are based on moments.

We performed those tests by taking the logarithm of the data and used the following functions in \verb|R|: \verb|shapiro.test|, \verb|sf.test|, \verb|lillie.test|, \verb|cvm.test|, \verb|ad.test|, \verb|dagoTest|, \verb|jarqueberaTest|, complemented with the \verb|R| package \verb|nortest| for the Shapiro--Francia, Lilliefors, Cramer--von Mises, and Anderson--Darling tests, and with the package \verb|fBasics| for the D'Agostino--Pearson and Jarque--Bera tests \citep{Gross2015,Wuertz2022}.

As a graphical validation of the lognormal model, those analytical tests were complemented with a Q--Q plot and a rank-size plot for each year. We performed the first one by taking the logarithm of the data and plotting them versus
quantiles from a standard normal distribution by using the functions \verb|qqnorm| and \verb|qqline| in \verb|R|. The second graph was created by plotting the theoretical lognormal survival function (obtained by maximum likelihood estimation) and the empirical survival function, both multiplied by $N+1$ and on a log-log scale.

\subsection{Gibrat's Law}\label{gibrat}

To provide additional support for the lognormal distribution, we propose to analyse whether Gibrat's law \citep{Gibrat1931} is verified, which implies that the growth rate of a stochastic process does not depend on its size but is proportionate to it, leading to the lognormal distribution for size. Note that Gibrat implies lognormal, but lognormal does not imply Gibrat.

By considering $S_{i,t}$ as the emissions of country $i$ at time $t$ and $S_{i,t-1},$ as the emissions of country $i$ in the previous period, we review the genesis of Gibrat's law adapted to the emissions variables. Thus, \cite{Gibrat1931} based its law on the assumption that in a certain time $t$, the change in variable $S$ is a random proportion of a function of the same variable in a previous time $t-1$; that is to say:
\begin{equation}
S_t-S_{t-1}=\epsilon_t g(S_{t-1})
\end{equation}
where $\epsilon_t$ are mutually independent and independent of $S_{t-1}$. By considering $g(S)\equiv S$, the result is \emph{the law of proportionate effect} assuming that $\epsilon_t$ is a sequence of independent and identically distributed (i.i.d.) random variables. Then, the change in $S$ at any step of the process is considered a random proportion of the previous value of $S$:
\begin{equation}
S_t-S_{t-1}=\epsilon_t S_{t-1}
\end{equation}
By operating and applying summation, the following equation is produced:
\begin{equation}\label{cociente}
\sum_{t=1}^{N}\frac{S_t-S_{t-1}}{S_{t-1}}=\sum_{t=1}^{N}\epsilon_t
\end{equation}
Thus, if we consider $S_{t-1}=S_{t-h}$ with $h \in \mathds{R}$, then the following results:
\begin{equation}
\sum_{t=1}^{N}\frac{S_t-S_{t-1}}{S_{t-1}}=\sum_{t=1}^{N}\frac{S_t-(S_{t-h})}{S_{t-h}}=\sum_{t=1}^{N}\frac{h}{S_{t-h}}
\end{equation}
It is necessary to assume that the effect at each step ($h$) is small:
\begin{equation}\label{aproximacion}
\sum_{t=1}^{N}\frac{h}{S_{t-h}}\approx \int_{S_0}^{S_N}\frac{1}{S}dS=\log(S_N)-\log(S_0)
\end{equation}
Notably, it is assumed that the change between each step is depreciable. From (\ref{cociente}) and (\ref{aproximacion}), it is possible to obtain:
\begin{equation}\label{final}
\log(S_N)=\log(S_0)+\sum_{t=1}^{N}\epsilon_t
\end{equation}
By assuming that $\epsilon_t$ are i.i.d., it is possible to conclude that $S_N$ are asymptotically lognormally distributed by using the Lindeberg-Levy Central Limit Theorem (CLT). This same conclusion can be obtained even if $\epsilon_t$ are heterogeneous, only using more general CLTs, \citep{Kleiber2003}. From this assumption, Gibrat's law \citep{Gibrat1931} was initially established in relation to the income of an individual (or the size of a firm). This law considers that the joint effect of a large number of mutually independent causes that have worked during a long period of time leads to a multiplicative random process in which the product of a large number of individual random variables tends to result in the lognormal distribution, \citep{Kleiber2003}.

Once the genesis of Gibrat's law was described, it was necessary to apply an adequate methodology for its verification. Within the field of emissions, \cite{Ahundjanov2019} stated that a simple method to empirically verify the law of proportionate effect consists of assuming from (\ref{final}) that the growth rate between any two periods can be described by a random walk process given by:
$$\log(S_{i,t})=\log(S_{i,t-1})+\zeta_{i,t}$$
where $\zeta_{i,t}=\alpha_i+v_{i,t}$, with $\alpha_i$ as the effect of individual factors on individual growth rates and $v_{i,t}$ as an i.i.d. distributed random effect. Thus, to verify Gibrat's law, this methodology suggests estimating the following linear regression:
\begin{equation}\label{m1}
\log(S_{i,t})=\alpha+\beta \log(S_{i,t-1})+v_{i,t}
\end{equation}
where the parameter $\alpha$ is interpreted as the common effect, while the parameter $\beta$ represents the country-specific growth rate. If the value of $\beta$ is equal to 1, the emissions growth rate is independent of the initial emissions. If $\beta<1$, smaller emissions grow faster than larger emissions, while if $\beta>1$, larger emissions grow faster than smaller emissions. To analyse whether Gibrat's law is verified, the null hypothesis $\beta=1$ is contrasted with \citep{Shapiroetal1987}, \citep{Clarcketal1992} and \citep{Chester1979}. If it is not rejected, the emissions growth rate and the initial emissions are independently distributed, and consequently, Gibrat's law is verified with a given level of confidence.

Additionally, the alternative specifications proposed by \cite{Eeckhout2004} can be adapted to verify the law of proportionate effect for CO$_2$ emissions from the following parametric regressions:
\begin{equation}\label{m2}
\frac{S_{i,t}}{S_{i,t-1}}=\alpha+\beta \frac{S_{i,t}+S_{i,t-1}}{2}+v_{i,t}
\end{equation}
\begin{equation}\label{m3}
\frac{S_{i,t}}{S_{i,t-1}}=\alpha+\beta S_{i,t-1}+v_{i,t}
\end{equation}
\begin{equation}\label{m4}
\log\left(\frac{S_{i,t}}{S_{i,t-1}}\right)=\alpha+\beta S_{i,t-1}+v_{i,t}
\end{equation}
In these cases, if the null hypothesis $\beta=0$ is not rejected, the coefficient on size will be insignificant for growth.

\cite{Ahundjanov2019} examined Gibrat's law of proportionate growth for national CO$_2$ emissions for 200 countries in five-year periods from 1995 to 2010, performing the specifications (\ref{m1}), (\ref{m2}), (\ref{m3}) and (\ref{m4}) for 2010-2005, 2005-2000 and 2000-1995. From all these methodologies, the authors concluded that there was strong statistical evidence in support of Gibrat's law. In this paper, we extended this analysis for the full period 1970-2021 by comparing a year to the immediately preceding year instead of comparing across five-year periods. We consider that with this more detailed analysis, it is easier to assume that the change between each step ($h$) is depreciable (as is established in expression \ref{aproximacion}). 
For a more general study of alternative methodologies, see \cite{Santarellietal2006}, where numerous empirical studies testing Gibrat's Law for firm size growth rate are reviewed.

\section{A climate policy tool for spatial allocation, at the country level, of a global fossil CO$_2$ emission goal}
\label{tool}

In this section, we propose a practical application developed from the previous theoretical basis. Thus, we describe how the findings in this paper could be used as a climate policy tool to convert a worldwide emission goal into national reduction targets.

We let $Y_1,Y_t$ denote the worldwide fossil CO$_2$ emissions in a base year (year 1) and in a target year (year $t$), respectively. Then, we define the ratio $R$ between those values $Y_1,Y_t$ as follows:
\begin{equation}\label{Eq.331}
R=Y_t/Y_1,
\end{equation}
This can be interpreted in terms of a reduction target of the global fossil CO$_2$ emissions between that year of interest and that base year.

With respect to the base year, we can obtain the value of $Y_1$ by adding the partial values of fossil CO$_2$ emitted by each country from our database.

With respect to the target year (year $t$), we let $X_t$ denote the amount of fossil CO$_2$  by a randomly selected country in that future year, and make the assumption that it follows a 2-parameter lognormal model in the whole range of the distribution (all world countries),
$X_t\sim \text{LN}(\mu_t,\sigma_t^2)$.
Then, we could opt to establish that the future amount of CO$_2$ emitted by each country should be in accordance with the rank of each country in the base year (from highest to lowest emissions), following that lognormal model. Finally, we could obtain those values by using the quantile function of the lognormal distribution. Based on that idea, the value of
$Y_t$ can be obtained by adding those values, and the ratio $R$ (\ref{Eq.331}) can be expressed as follows:
\begin{equation}\label{Eq.332}
R=
\displaystyle\frac
{\displaystyle\sum_{i=1}^{N}F^{-1}\left(\displaystyle\frac{i}{N+1}\right)}
{\displaystyle\sum_{i=1}^{N}x_{i1}}
=
\displaystyle\frac
{\displaystyle\sum_{i=1}^{N}\exp\left[\mu_t+\sigma_t\Phi^{-1}
\left(\displaystyle\frac{i}{N+1}\right)\right]}
{\displaystyle\sum_{i=1}^{N}x_{i1}}
\end{equation}
where $x_{11}\leq\dots\leq x_{i1}\leq\dots\leq x_{N1}$ is the amount of fossil CO$_2$ emitted by each of the $N$ countries during the base year (the rank of that country is given by $rank=N+1-i$), where we assume that the number of countries does not change in both years, $\Phi$ is the cumulative distribution function (CDF) of the standard normal distribution \citep{Kleiber2003}, and $F$ is, in this case, the CDF of the 2-parameter lognormal model.

Formula (\ref{Eq.332}) gives us one degree of freedom,
$\mu_t$ or $\sigma_t$ to reach a specific reduction target $R$ of global fossil CO$_2$ emissions.

To interpret the parameter $\mu$ in a log-normally distributed random variable $X\sim \text{LN}(\mu,\sigma^2)$, the parameter $\mu$ is the mean (and the median) of that random variable's logarithm $\log(X)$, and in addition, the parameter $\lambda=e^\mu$ is the scale parameter of the random variable $X$
\citep{Sarabia2007}.

To illustrate this, let us consider two years in which the amount of fossil CO$_2$ emitted by a randomly selected country ($X_1$ and $X_t$, respectively) was log-normally distributed. In addition, let us consider that there was no change in the parameter $\sigma$. In that case, we have $X_1=e^{\mu_1+\sigma Z}$ and $X_t=e^{\mu_t+\sigma Z}$, where $Z \sim \text{N}(0,1)$ is a standard normal random variable. Then, we obtain that $X_t/X_1=e^{\mu_t-\mu_1}=e^{\Delta \mu}$. Therefore, the term $e^{\Delta\mu}$ could be interpreted as the change in scale of the fossil CO$_2$ emissions of the countries between a base year and a target year.

To interpret parameter $\sigma$, two widely used inequality measures, the Theil entropy index and the mean log deviation (MLD), coincide in the case of the two-parameter lognormal model:
$E[(X/\mu)\log(X/\mu)]=-E[\log(X/\mu)]=\sigma^2/2$ \citep{Sarabia2017}. Again we consider two years in which the amount of fossil CO$_2$ emitted by a randomly selected country ($X_1$ and $X_t$, respectively) was log-normally distributed. Both inequality measures are denoted as $T$. We have that
$\Delta T = T_t - T_1 = \sigma_t^2/2-\sigma_1^2/2$ corresponds to the evolution of the inequality in fossil CO$_2$ emissions at the country level, from a base year to a target year.

In summary, the expression (\ref{Eq.332}) tells us that a specific reduction target $R$ of the global fossil CO$_2$ emissions could be reached as a result of our decision in two dimensions: the change in scale and/or the change in inequality in fossil CO$_2$ emissions at the country level.

Finally, once we have made those decisions regarding scale and inequality, we can obtain the national reduction targets $r_i$ for each country. For that, we can consider a year of reference (year 2) that can be the same or not as the year of reference (year 1) used for the global target $R$---for example, we could take the last year our data are available---and therefore, analyse different scenarios as follows:
\begin{equation}\label{Eq.333}
r_i=\displaystyle\frac
{\displaystyle\exp\left[\mu_t+\sigma_t\Phi^{-1}
\left(\displaystyle\frac{i}{N+1}\right)\right]}
{x_{i2}},\;i=1,\dots,N.
\end{equation}
where $x_{12}\leq\dots\leq x_{i2}\leq\dots\leq x_{N2}$ is the amount of fossil CO$_2$ emitted by each of the $N$ countries during that year of reference (year 2), and where we denote those national targets in lowercase to distinguish them from the global target $R$ in uppercase.
\newpage

\section{Results and discussion}\label{results}

\subsection{Modelling the whole range of the distribution (all world countries)}

\begin{figure}[p]\centering
\includegraphics[width=1\textwidth]{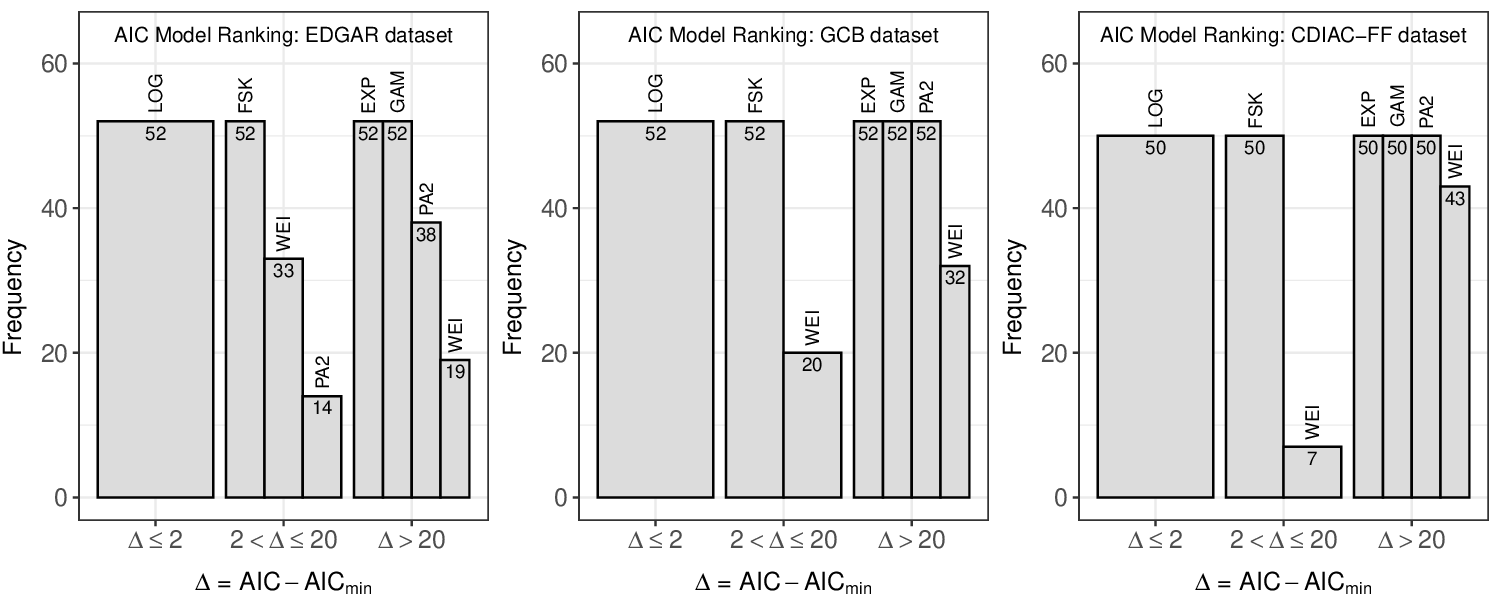}
\caption{AIC rankings:  lognormal (LOG), exponential (EXP), Fisk (FSK), gamma (GAM), Lomax (PA2) and Weibull (WEI) models. Number of years (frequency) for each of the models that best fits the data ($\Delta=AIC-AIC_{min}\leq 2$), that has relatively little support ($2<\Delta\leq 20$), or that has no empirical support ($\Delta>20$), in the period 1979-2021 (52 years in total) for the EDGAR and GCB datasets, and in the period 1979-2019 (50 years in total) for the CDIAC-FF dataset.}
\label{fig:1}
\end{figure}

Figure \ref{fig:1} compares the number of years (frequency) for each of those models that best fits the data ($\Delta=AIC-AIC_{min} \leq 2$), that has relatively little support ($2<\Delta\leq 20$), or that has no empirical support ($\Delta>20$). 
That figure \ref{fig:1} shows the AIC ranking among the lognormal (LOG), exponential (EXP), Fisk (FSK), gamma (GAM), Lomax (PA2) and Weibull (WEI) models.

On the one hand, it can be noted that the lognormal model seems to better fit the data in all three datasets considered (EDGAR, GCB and CDIAC-FF datasets) and in all the years considered (52 years in the EDGAR and GCB datasets and 50 years in the CDIAC-FF dataset), compared to the other five models considered. On the other hand, in all the datasets and years considered, the Fisk model has relatively little support; the Weibull, Lomax models have relatively little support in some cases and no empirical support in the rest of them; and the exponential, gamma models have no empirical support. 

Therefore, we found that the 2-parameter lognormal model was the best model in comparison with the other five models considered - confirming that a comparison based directly on $AIC$, $BIC$ or $HQC$ statistics (Akaike, Bayesian or Hannan--Quinn Information Criterion) gave the same result in favour of the 2-parameter lognormal model.

\begin{figure}[p]\centering
\includegraphics[width=0.82\textwidth]{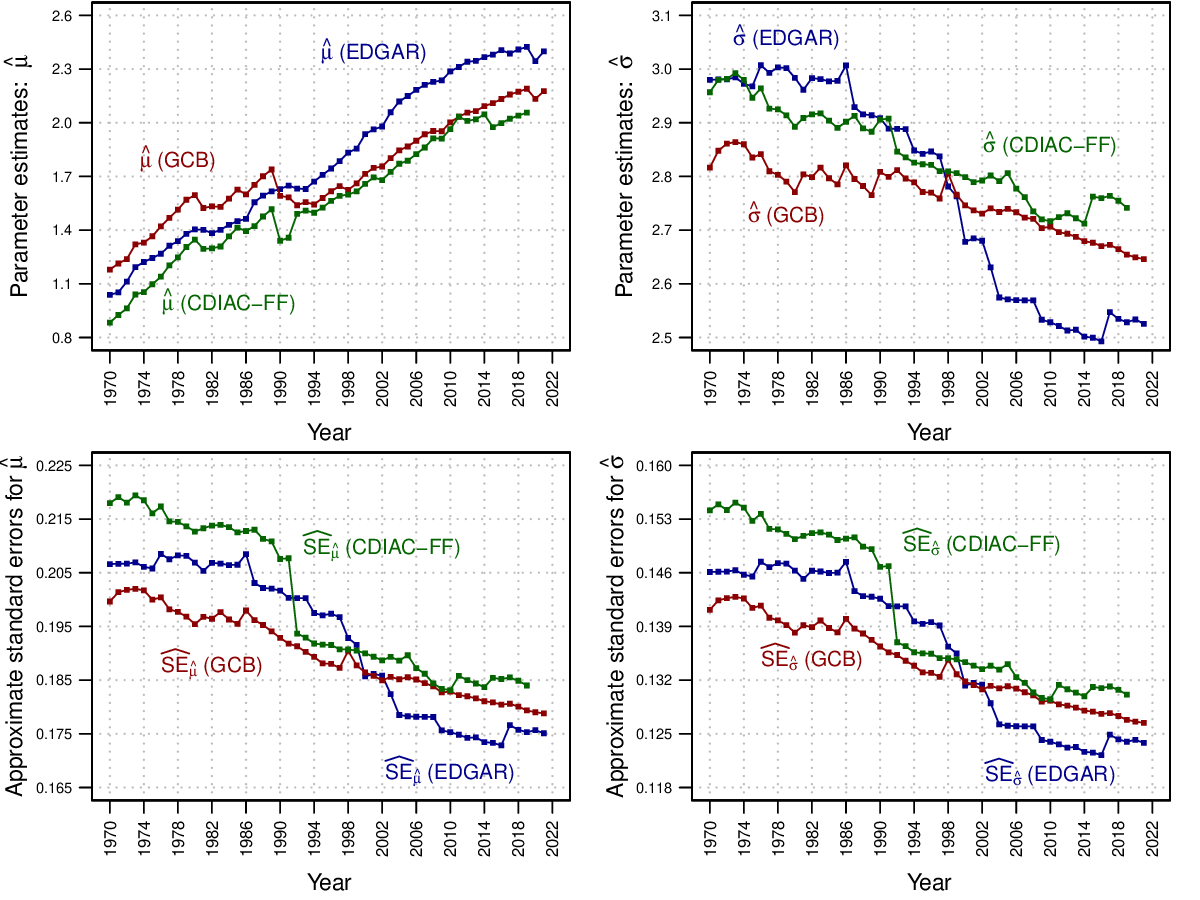}
\caption{
Parameter estimates for the lognormal distribution, fitted by maximum-likelihood estimation (top part), and the corresponding standard errors (bottom part), for each dataset and year.}
\label{fig:2}
\end{figure}

Figure \ref{fig:2} shows the parameter estimates $(\hat{\mu},\hat{\sigma})$ for the lognormal model, fitted by maximum-likelihood estimation (top part), and the corresponding standard errors 
$(\widehat{SE}_{\hat{\mu}},\widehat{SE}_{\hat{\sigma}})$ (bottom part), for each dataset and year considered (the EDGAR, GCB, CDIAC-FF datasets, and the periods 1970-2021, 1970-2021 and 1970-2019, respectively). The trend of the parameter estimates $\hat{\mu}$ presents an increasing tendency, and the parameter estimate for $\hat{\sigma}$ presents a decreasing tendency over the
years.\footnote{Figure \ref{fig:2} shows the similarity between both plots of the standard errors $(\widehat{SE}_{\hat{\mu}},\widehat{SE}_{\hat{\sigma}})$, in line with the Fisher information of the lognormal distribution with a vector of parameters
${\bm\theta}=(\mu,\sigma^2)$ in one observation \cite{Kleiber2003}:
$I({\bm\theta})=\left[ {\begin{array}{cc}
    1/\sigma^2 & 0 \\
    0 & 1/(2\sigma^4) \\
  \end{array} } \right]$, which only depends on $\sigma$ and not on $\mu$.}

\begin{figure}[p]\centering
\includegraphics[width=1\textwidth]{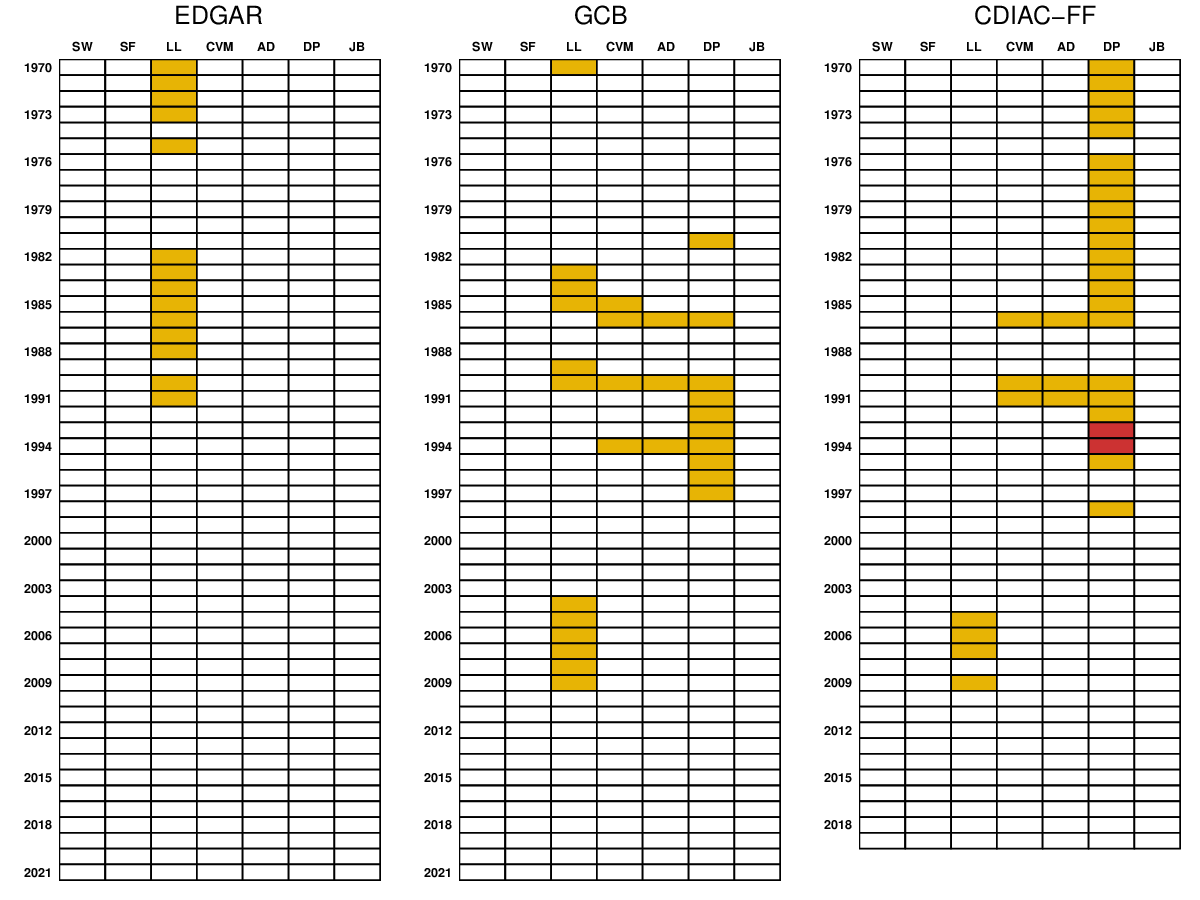}
\caption{$p-\text{value}>0.05$ (white cells), $0.01<p-\text{value}<0.05$ (yellow cells), and $p-\text{value}<0.01$ (red cells), obtained for the three datasets (EDGAR, GCB and CDIAC-FF) and the period 1970-2021 considered, by using Shapiro-Wilk (SW), Shapiro-Francia (SF), Lilliefors (LL), Cramer-von Mises (CVM), Anderson-Darling (AD),  D'Agostino-Pearson Omnibus (DP) and Jarque-Bera (JB) tests.}
\label{fig:3}
\end{figure}

Tables \ref{tab2} and \ref{tab3} show the number of years in which the lognormal model can be ruled out with significance levels of 0.05 and 0.01, respectively, in the three datasets considered: the first for the period 1970-2021 and the second for recent years (2000-2021). Figure \ref{fig:3} shows the $p$-values obtained greater than 0.05 (cells coloured in white), the $p$-values obtained that are less than 0.05 and greater than 0.01 (coloured in yellow), and the $p$-values obtained that are less than 0.01 (in red).
It can be noted that
\begin{enumerate}[leftmargin=5 mm,nolistsep]
\item The lognormal model cannot be ruled out with a significance level of 0.01 in the three datasets and in all 52, 52, and 50 years considered, with two exceptions:  years 1993 ($p$-value=0.008) and 1994 ($p$-value=0.009), in the CDIAC-FF dataset and with the D'Agostino-Pearson test (with both $p$-values very close to 0.01);
\item The lognormal model cannot be ruled out with a significance level of 0.05 by using Shapiro--Wilk, Shapiro--Francia, and Jarque--Bera tests in the three datasets and in all the years considered;
\item The lognormal model can be ruled out with a significance level of 0.05 in many years
by using Lilliefors, Cramer--von Mises, Anderson--Darling, and D'Agostino-Pearson tests. In no year, however, can the lognormal model be ruled out considering the three datasets all at once with either of those four tests;
\item Considering recent years, the period 2000-2021 (22 years in the EDGAR and CCB and 20 years in the CDIAC-FF), the lognormal model cannot be ruled out with a significance level of 0.05, with the exception of 6 years in the GCB and 4 years in the CDIAC-FF datasets, by using the Lilliefors test (see Table \ref{tab3}).
\end{enumerate}

\begin{table}[htp]\footnotesize
\renewcommand{\tablename}{\footnotesize{Table}}
\caption{\footnotesize\label{tab2}Period: 1970-2021. Data: EDGAR and GCB (52 years), CDIAC-FF (50 years). Number of years that the lognormal model can be ruled out with a significance level of 0.05 \& 0.01.}
\centering
\setlength{\tabcolsep}{1.8 mm}
\begin{tabular}{@{}l @{\hspace{0.8cm}} c c c @{\hspace{0.8cm}} c c c @{}}
\toprule
Dataset&EDGAR& GCB& CDIAC-FF&EDGAR& GCB& CDIAC-FF\\
Num.Years&(52)&(52)&(50)&(52)&(52)&(50)\\
Significance level &\multicolumn{3}{c}{$\alpha=0.05$}
&\multicolumn{3}{c}{$\alpha=0.01$}\\
\midrule
Shapiro--Wilk  & 0 & 0 & 0 & 0 & 0 & 0\\
Shapiro--Francia & 0 & 0 & 0 & 0 & 0 & 0\\
Lilliefors  & \bf{14} & \bf{12} & \bf{4} & 0 & 0 & 0\\
Cramer--von Mises  & 0 & \bf{4} & \bf{3} & 0 & 0 & 0\\
Anderson--Darling  & 0 & \bf{3} & \bf{3} & 0 & 0 & 0\\
D'Agostino-Pearson  & 0 & \bf{10} & \bf{23} & 0 & 0 & \bf{2}\\
Jarque--Bera  & 0 & 0 & 0 & 0 & 0 & 0\\
\bottomrule
\end{tabular}
\end{table}

\begin{table}[p]\footnotesize
\renewcommand{\tablename}{\footnotesize{Table}}
\caption{\footnotesize\label{tab3}Period: 2000-2021. Data: EDGAR and GCB (22 years), CDIAC-FF (20 years). Number of years that the lognormal model can be ruled out with a significance level of 0.05 \& 0.01.}
\centering
\setlength{\tabcolsep}{1.8 mm}
\begin{tabular}{@{}l @{\hspace{0.8cm}} c c c @{\hspace{0.8cm}} c c c @{}}
\toprule
Dataset&EDGAR& GCB& CDIAC-FF&EDGAR& GCB& CDIAC-FF\\
Num.Years&(22)&(22)&(20)&(22)&(22)&(20)\\
Significance level &\multicolumn{3}{c}{$\alpha=0.05$}
&\multicolumn{3}{c}{$\alpha=0.01$}\\
\midrule
Shapiro--Wilk  & 0 & 0 & 0 & 0 & 0 & 0\\
Shapiro--Francia & 0 & 0 & 0 & 0 & 0 & 0\\
Lilliefors  & 0 & \bf{6} & \bf{4} & 0 & 0 & 0\\
Cramer--von Mises  & 0 & 0 & 0 & 0 & 0 & 0\\
Anderson--Darling  & 0 & 0 & 0 & 0 & 0 & 0\\
D'Agostino-Pearson  & 0 & 0 & 0 & 0 & 0 & 0\\
Jarque--Bera  & 0 & 0 & 0 & 0 & 0 & 0\\
\bottomrule
\end{tabular}
\end{table}

\begin{figure}[ht]\centering
\includegraphics[width=0.75\textwidth]{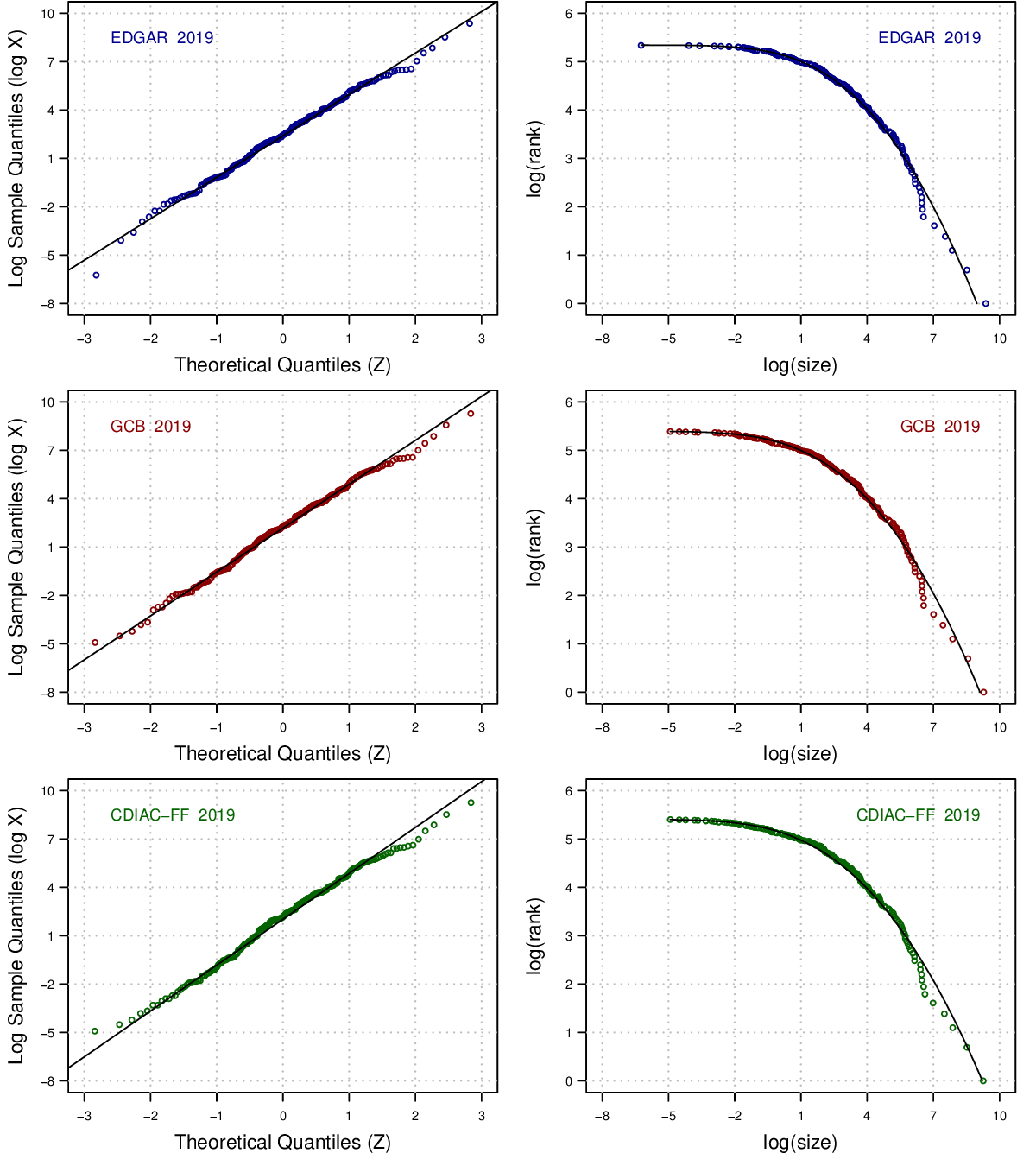}
\caption{Q-Q plots (left) and rank-size plot (right) for the EDGAR, GCB and CDIAC-FF datasets for 2019.}
\label{fig:4}
\end{figure}

Figure \ref{fig:4} shows the Q-Q plots (left) and rank-size plots (right) for the EDGAR, GCB and CDIAC-FF datasets for the year 2019 (the last year in common in those three datasets). The lognormal model fit the data reasonably well in that year.

In summary, the results indicate that the data are consistent with the lognormal hypothesis at the 0.01 level of significance and that the lognormal model
cannot be firmly ruled out at the 0.05 level of significance, especially in recent years.
Therefore, we found that the lognormal model can be useful for describing
the fossil CO$_2$ emissions data, at the country level, in the whole range of the distribution (all world countries).
\subsection{Gibrat's Law}

Table \ref{tab4} summarizes the different methodologies that will be applied to analyse Gibrat's law of proportionate growth for national CO$_2$ emissions, which are presented in subsection \ref{gibrat}. Note that for the first methodology (M1), the null hypothesis is $\beta=1$, while the null hypothesis of the rest of the methodologies (M2, M3 and M4) is $\beta=0$.

\begin{table}[p]\footnotesize
\renewcommand{\tablename}{\footnotesize{Table}}
\caption{\footnotesize\label{tab4}Summary of the methodologies applied to analyse Gibrat's law of proportionate growth for national CO$_2$ emissions.}
\centering
\setlength{\tabcolsep}{2 mm}
\begin{tabular}{@{}l c c @{}}
\toprule		
Methodology  & Regression  & $H_0$ \\
\midrule
M1 & $\log(S_{i,t})=\alpha+\beta \log(S_{i,t-1})+v_{i,t}$ & $\beta=1$\\[0.5ex]
M2 & $\frac{S_{i,t}}{S_{i,t-1}}=\alpha+\beta \frac{S_{i,t}+S_{i,t-1}}{2}+v_{i,t}$ & $\beta=0$\\[0.5ex]
M3 & $\frac{S_{i,t}}{S_{i,t-1}}=\alpha+\beta S_{i,t-1}+v_{i,t}$ & $\beta=0$\\[0.5ex]
M4 & $\log\left(\frac{S_{i,t}}{S_{i,t-1}}\right)=\alpha+\beta S_{i,t-1}+v_{i,t}$ & $\beta=0$\\
\bottomrule
\end{tabular}
\end{table}

\begin{figure}[htp]\centering
\includegraphics[width=1\textwidth]{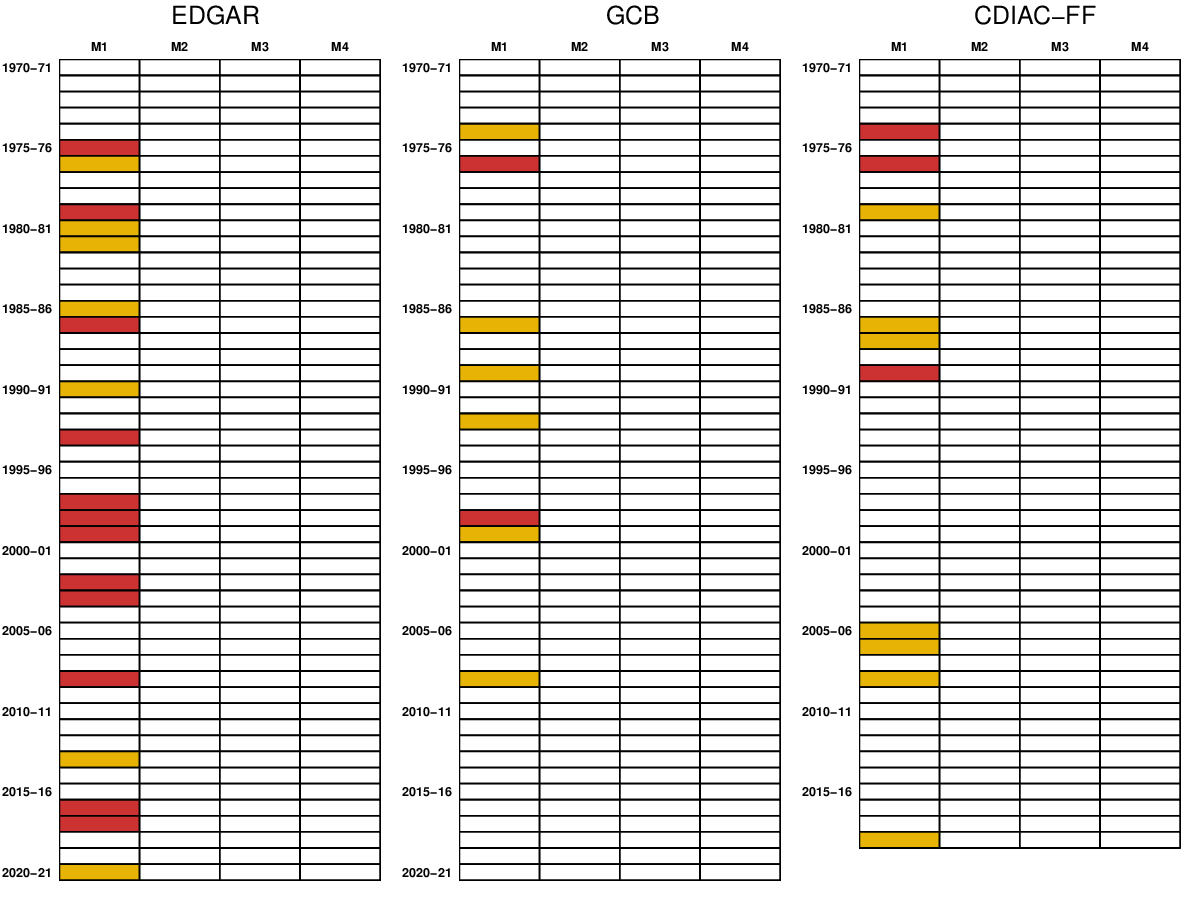}
\caption{$p-\text{value}>0.05$ (white cells), $0.01<p-\text{value}<0.05$ (yellow cells), and $p-\text{value}<0.01$ (red cells), obtained for the three datasets (EDGAR, GCB and CDIAC-FF) and the period 1970-2021 considered by methodologies M1, M2, M3 and M4 summarized in Table \ref{tab4}}
\label{fig:5}
\end{figure}

Figure \ref{fig:5} shows the $p$-values obtained that are greater than 0.05 (cells coloured in white), the $p$-values obtained that are less than 0.05 and greater than 0.01 (coloured in yellow), and the $p$-values obtained that are less than 0.01 (in red). Note that for methodologies M2, M3 and M4, the null hypothesis ($\beta=0$) cannot be rejected for any level of confidence (90\%, 95\% and 99\%), while in the case of methodology M1, the null hypothesis ($\beta=1$) is rejected in some years that do not always coincide between the different databases.

With the aim of deepening these results, Table \ref{tabA2} shows the value of beta for the different methodologies (M1, M2, M3 and M4) and databases. Note that for methodologies M2, M3 and M4, the value of beta is practically zero (up to the fifth decimal place) for all the years studied.

In the case of methodology M1, the range of $\beta$ is (0.97-1.02), (098-1.01) and (0.98-1) for the Edgar, GCB and CDIAD-FF databases, respectively. Rounding to three decimal places, the average is equal to 0.996 in the three databases, while the standard deviation is equal to 0.009, 0.005 and 0.005 for the Edgar, GCB and CDIAD-FF databases, respectively. We conclude that Gibrat's law is verified with strong statistical evidence since the null hypothesis is rejected for only one of the methodologies (M1), and even when it is rejected, the value of $\beta$ is very close to one.

This conclusion is in line with the results obtained by \cite{Ahundjanov2019}, who examined Gibrat's law of proportionate growth for national CO$_2$ emissions for 200 countries obtained from methodology M1:
\begin{enumerate}[leftmargin=5mm,nolistsep]
\item The null hypothesis ($\beta=1$) was rejected at the 90\% and 95\% levels of confidence but not at a 99\% level in the three analysed periods (2010-2005, 2005-2000 and 2000-1995),
\item while the null hypothesis ($\beta=0$) was not rejected for the rest of the applied methodologies (M2, M3 and M4) in the three analysed periods.
\end{enumerate}
Our extension to analyse the entire period of 1970-2020, comparing one year with the previous year, has resulted in a deeper analysis of the verification of Gibrat's law.

\subsection{Calculating countries' fossil CO$_2$ emissions targets}
In this section, we illustrate with an example how the climate policy tool described in subsection (\ref{tool}) can be used to convert a global emission goal into national reduction targets.

For this example, we consider as a reference the European Climate Law \cite{EU2030}, which establishes an intermediate global emission target for the year 2030 of reducing net greenhouse gas emissions by at least 55\% compared to the year 1990. We set that goal for worldwide fossil CO$_2$ emissions in particular. In addition, we explain this example by using the EDGAR v7.0 database described in section \ref{data}, with information from $N=208$ countries in the base year 1990. According to that target, the ratio $R$ (reduction target of the global fossil CO$_2$ emissions between 1990 and 2030) should take the value $R=(1-0.55)=0.45$ at most.

Once the lognormal distribution is selected as an appropriate model for describing global fossil CO$_2$ emissions (see two previous sections), it is also possible to analyse the parameters of this distribution as a function of time. Table \ref{tab5} shows the estimation of the linear relation of $\mu$ and $\sigma$ as a function of the year for the EDGAR database (period 1970-2021). Notably, for $\mu$, the relation is positive, while the relation is negative in the case of $\sigma$ (in concordance with figure \ref{fig:2}). In both cases, the linear regressions present a high goodness of fit ($R$-squared higher than 0.9) having both global significance and all parameters individually significant (both tests with a level of confidence of 99\%). Possible autocorrelation was corrected by using robust estimation methodologies. Because the goal of this research is to make predictions, multicollinearity was not analysed.

By using the previous models, Table \ref{tab6} displays the prediction of
$\mu_t$ and $\sigma_t$ for 2025, 2030 and 2035 for the EDGAR database.
In addition, and as a reference, it shows the corresponding values of the global ratio $R$ obtained from those predictions, obtained by using the expression (\ref{Eq.332}). It can be noted that for year 2030, the global ratio $R$ takes the value 1.6180, which is greater than 1 and very far from the reduction target $R=0.45$ for that year 2030 described above.

Then, to reach the global target $R=0.45$ for this example, we could opt to keep the inequality estimates (given by $\sigma_t=2.3474$) for 2030 and calculate the scale needed (given by $\mu_t$) to obtain our global emission goal $R=0.45$ under that scenario. In that case, by using expression (\ref{Eq.332}) and by using the \verb|R| software function \verb|uniroot| (within the \verb|rootSolve| \verb|R| package), we obtain $\mu_t=1.5053$.

Finally, substituting those values into the expression (\ref{Eq.333}), and for this example, considering the same year of reference for national reduction targets (year 2) and for the global target (year 1 = 1990), we have

\begin{equation}\label{Eq.334}
r_i=\displaystyle\frac
{\displaystyle\exp\left[1.5053+2.3474\;\Phi^{-1}
\left(\displaystyle\frac{i}{208+1}\right)\right]}
{x_{i1}},\;i=1,\dots,208.
\end{equation}

\begin{figure}[p]\centering
\includegraphics[width=1\textwidth]{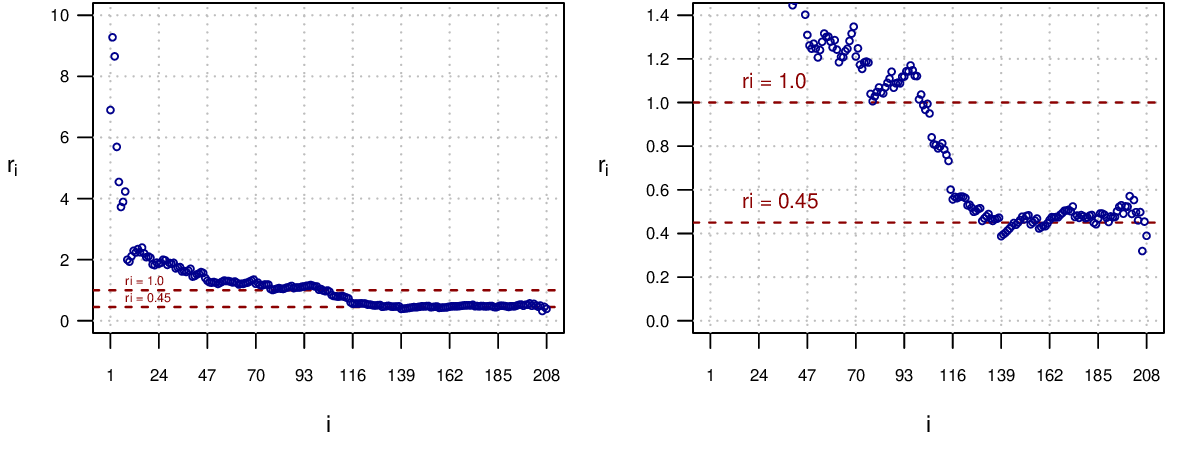}
\caption{National reduction target $r_i$, as obtained by using the Edgar dataset from expression (\ref{Eq.334}) based on a global target $R=0.45$ and the inequality estimates for 2030 (given by $\sigma_t=2.3474$) for all 208 countries (left), and a more detailed view around the value $r_i=0.45$ (right)}
\label{fig:6}
\end{figure}

Figure \ref{fig:6} shows the value of $r_i$ as obtained by using the previous expression (\ref{Eq.334}) (on the left, for all 208 countries from the EDGAR dataset, and on the right, a more detailed view around the value $r_i=0.45$). Notably, in this example, we have three groups of countries:
\begin{enumerate}[leftmargin=5mm,nolistsep]
\item low-emission countries---with a value of $r_i$ greater than 1, which means that they could increase their levels of fossil CO$_2$ emissions or sell it to others;
\item middle-emission countries---with a value of $r_i$ between 0.45 and 1, that should reduce their emissions, but not as much as the global target $R=0.45$;
\item high-emission countries---with a value of $r_i$ close to or less than 0.45.
\end{enumerate}

\begin{table}[p]\footnotesize
\renewcommand{\tablename}{\footnotesize{Table}}
\caption{\footnotesize\label{tab5}Estimation of the linear relation between $\mu$ and $\sigma$ as a function of the time (year), using EDGAR v7.0 database (1970-2021): $\mu=\alpha+\beta year+u$ and $\sigma=\alpha+\beta year+u$ respectively.}
\centering
\setlength{\tabcolsep}{4 mm}
\begin{tabular}{@{}l d{3.2} d{3.2} @{}}
\toprule
    &   \multicolumn{1}{c}{$\mu=\alpha+\beta year+u$}
        &   \multicolumn{1}{c}{$\sigma=\alpha+\beta year+u$}\\
\midrule
$\hat{\alpha}$                 & -55.3221 ***   & 27.5025 *** \\
Std. Error$(\hat{\alpha})$   & (1.0647)       & (1.1335)     \\[0.5ex]
$\hat{\beta}$                   & 0.0286 ***    & -0.0124 ***   \\
Std. Error$(\hat{\beta})$     & (0.0005)      & (0.0006)       \\[0.5ex]
$R$-squared                           & 0.9829         & 0.9049         \\
$p$-value $F$                & <0.0001  & <0.0001  \\
\bottomrule
\end{tabular}
\end{table}

\begin{table}[p]\footnotesize
\renewcommand{\tablename}{\footnotesize{Table}}
\caption{\footnotesize\label{tab6}Prediction for $\mu_t$ and $\sigma_t$, from models estimated in table \ref{tab5}, for years 2025, 2030 and 2035 (Edgar database), and the corresponding value of the global ratio $R$.}
\centering
\setlength{\tabcolsep}{7 mm}
\begin{tabular}{@{} l c  c c @{}}
\toprule
Year &$\hat{\mu}_t$&$\hat{\sigma}_t$&R\\[0.5ex]
\midrule
2025  & 2.6418   & 2.4094   & 1.5693\\
2030  & 2.7850   & 2.3474   & 1.6180\\
2035  & 2.9281   & 2.2854   & 1.6711\\
\bottomrule
\end{tabular}
\end{table}

\section{Conclusion}
\label{conclusion}

Through rigorous empirical analysis, this paper shows the adequacy of the lognormal distribution in characterizing global fossil CO$_2$ emissions for the three datasets analysed and all world countries, bolstered by the validation of Gibrat's law with strong statistical evidence. Additionally, we offer projections for lognormal distribution parameters for the years 2025, 2030, and 2035.

Taking advantage of the fact that CO$_2$ emissions have been well adjusted with a simple distribution such as the lognormal distribution (with only two easy-to-interpret parameters), this paper theoretically develops a tool to effectively translate worldwide emission goals into targeted national reduction targets.

After setting a reference value in the target established in the European Climate Law for net greenhouse gas emissions, the usefulness of the tool is illustrated with the EDGAR v7.0 database and the target of reducing worldwide fossil CO$_2$ emissions by at least
55\% for the year 2030 compared to the year 1990. We find that the global target is redistributed to the different countries, finding three groups of countries:
\begin{enumerate}[leftmargin=5mm,nolistsep]
\item Low-emission countries that could increase their levels of fossil CO$_2$ emissions or sell it to others;
\item Middle-emission countries that should reduce their emissions but not as much as the global target;
\item And high-emission countries that should reduce their emissions even slightly above the global target.
\end{enumerate}

Future research could delve into the proposed tool to analyse how the results are related to the economic development of the countries and other factors that may influence emissions and their implications for climate change. Furthermore, this tool could be applied in other research fields.

\section*{Funding}
This work was partially funded (F.P.) by grant no. PID2019-105986GB-C22 by MCIN/AEI/10.13039/501100011033 (Ministerio de Ciencia e Innovaci\'on, Spain).

\section*{Acknowledgments}
We acknowledge: the European Commission, Joint Research Centre (JRC), and the International Energy Agency (IEA) for the EDGAR v7.0 database; the Global Carbon Project, for the GCB 2022 database; and the Research Institute for Environment, Energy, and Economics, Appalachian State University, for the CDIAC-FF 2022 database. 

\newpage
\appendix
\setcounter{table}{0}
\renewcommand{\thetable}{A\arabic{table}}

\section{Additional tables and figures}
%
%
\begin{table}[htp]\scriptsize
\renewcommand{\tablename}{\footnotesize{Table}}
\caption{\label{tabA1}\footnotesize Cumulative distribution functions $F(x)$ (CDF), probability density functions $f(x)$, support and parameters of the models selected for fitting the whole range of the data. $\Phi(z)$ denotes the standard normal CDF and $\gamma(a,z)$ denotes the lower incomplete gamma function
$\gamma(a,z)=\int_{0}^{z} t^{a-1} e^{-t} dt$.}
\centering
\setlength{\tabcolsep}{2 mm}
\begin{tabular}{@{}l c c c @{}}
\toprule		
Distribution  & $F(x)$  & $f(x)$  & Support; Parameters \\
\midrule
Exponential (EXP)
& $1-\exp\left(-\displaystyle\frac{x}{\sigma}\right)$
& $\displaystyle\frac{1}{\sigma}\exp\left(-\displaystyle\frac{x}{\sigma}\right)$
&$x>0;\sigma>0$\\[3ex]
Fisk (FSK)
& $\displaystyle\frac{1}{1+(x/\sigma)^{-\beta}}$
& $\displaystyle\frac{(\beta/\sigma)(x/\sigma)^{\beta-1}}{[1+(x/\sigma)^{\beta}]^2}$
&$x>0;\beta,\sigma>0$\\[3ex]
Gamma (GAM)
& $\displaystyle\frac{1}{\Gamma(\beta)}\gamma\left(\beta,\displaystyle\frac{x}{\sigma}\right)$
& $\displaystyle\frac{1}{\sigma\Gamma(\beta)}\left(\displaystyle\frac{x}{\sigma}\right)^{\beta-1}\exp\left(-\displaystyle\frac{x}{\sigma}\right)$
&$x>0;\beta,\sigma>0$\\[3ex]
Lognormal (LOG)
& $\Phi\left(\displaystyle\frac{\log x-\mu}{\sigma}\right)$
& $\displaystyle\frac{1}{x\sigma\sqrt{2\pi}}\exp\left[-\frac{(\log x-\mu)^2}{2\sigma^2}\right]$
&$x>0;\mu\in(-\infty,+\infty),\sigma>0$\\[3ex]
Lomax (PA2)
& $1-\left(1+\displaystyle\frac{x}{\sigma}\right)^{-\alpha}$
& $\displaystyle\frac{\alpha \sigma^\alpha}{(x+\sigma)^{\alpha+1}}$	
&$x>0;\alpha,\sigma>0$\\[3ex]
Weibull (WEI)
& $1-\exp\left[-\left(\displaystyle\frac{x}{\sigma}\right)^\alpha\right]$
& $\left(\displaystyle\frac{\alpha}{\sigma}\right)\left(\displaystyle\frac{x}{\sigma}\right)^{\alpha-1}\exp\left[-\left(\displaystyle\frac{x}{\sigma}\right)^\alpha\right]$
&$x>0;\alpha,\sigma>0$\\
\bottomrule
\end{tabular}
\end{table}
%
%
\begin{table}[h]\scriptsize
\renewcommand{\tablename}{\footnotesize{Table}}
\caption{\footnotesize\label{tabA2}Value of $\beta$ obtained from the three datasets (EDGAR, GCB and CDIAC-FF) and the period 1970-2021 by methodologies M1, M2, M3 and M4, as summarized in Table \ref{tab4}}
\centering
\setlength{\tabcolsep}{1.4 mm}
\begin{tabular}{@{}l @{\hspace{0.5cm}}
c c c c@{\hspace{0.5cm}} c c c c@{\hspace{0.5cm}}c c c c@{}}
\toprule
&\multicolumn{4}{c}{EDGAR}&
\multicolumn{4}{c}{GCB}&\multicolumn{4}{c}{CDIAC-FF}\\
$t-1$-$t$&M1&M2&M3&M4&M1&M2&M3&M4&M1&M2&M3&M4\\
\midrule
1970-71 & 1.00 & -1e-05 & -1e-05 & -1e-05 & 1.00 & -4e-05 & -5e-05 & -2e-05 & 0.99 & -5e-05 & -5e-05 & -2e-05 \\
1971-72 & 1.00 & -9e-06 & -9e-06 & -7e-06 & 1.00 & -9e-06 & -1e-05 & -4e-06 & 1.00 & -2e-05 & -2e-05 & -1e-05 \\
1972-73 & 1.00 & -2e-05 & -2e-05 & -1e-05 & 1.00 & -3e-05 & -3e-05 & -2e-05 & 1.00 & -3e-05 & -3e-05 & -2e-05 \\
1973-74 & 1.00 & -2e-05 & -2e-05 & -2e-05 & 1.00 & -1e-05 & -1e-05 & -7e-06 & 0.99 & -1e-05 & -1e-05 & -9e-06 \\
1974-75 & 1.00 & -2e-05 & -2e-05 & -2e-05 & 0.99 & -3e-05 & -3e-05 & -2e-05 & 0.99 & -3e-05 & -3e-05 & -2e-05 \\
1975-76 & 1.01 & 2e-05 & 2e-05 & 2e-05 & 1.00 & -1e-05 & -2e-05 & -7e-07 & 1.00 & -6e-06 & -7e-06 & 5e-06 \\
1976-77 & 0.99 & -1e-05 & -1e-05 & -1e-05 & 0.99 & -2e-05 & -3e-05 & -1e-05 & 0.99 & -3e-05 & -3e-05 & -2e-05 \\
1977-78 & 1.00 & -2e-06 & -3e-06 & -1e-06 & 1.00 & -2e-05 & -2e-05 & -9e-06 & 1.00 & -1e-05 & -1e-05 & -4e-06 \\
1979-79 & 1.00 & -9e-06 & -9e-06 & -7e-06 & 0.99 & -3e-05 & -3e-05 & -2e-05 & 0.99 & -3e-05 & -3e-05 & -2e-05 \\
1979-80 & 0.99 & -2e-05 & -2e-05 & -2e-05 & 0.99 & -4e-05 & -4e-05 & -2e-05 & 0.99 & -5e-05 & -5e-05 & -2e-05 \\
1980-81 & 0.99 & -1e-05 & -2e-05 & -9e-06 & 0.99 & -1e-05 & -1e-05 & -4e-06 & 0.99 & -1e-05 & -2e-05 & -4e-06 \\
1981-82 & 1.01 & 9e-07 & 3e-07 & 3e-06 & 1.00 & -2e-05 & -2e-05 & -1e-05 & 1.00 & -2e-05 & -2e-05 & -5e-06 \\
1982-83 & 1.00 & -1e-05 & -1e-05 & -7e-06 & 1.00 & -2e-06 & -3e-06 & 5e-06 & 1.00 & -6e-06 & -7e-06 & -5e-07 \\
1983-84 & 1.00 & -5e-06 & -6e-06 & 1e-06 & 0.99 & -2e-05 & -2e-05 & -8e-06 & 1.00 & -1e-05 & -1e-05 & -4e-06 \\
1984-85 & 1.00 & -8e-06 & -8e-06 & -6e-06 & 1.00 & -1e-05 & -2e-05 & -1e-05 & 0.99 & -1e-05 & -1e-05 & -7e-06 \\
1985-86 & 1.01 & -3e-06 & -4e-06 & 3e-06 & 1.01 & 9e-06 & 9e-06 & 2e-05 & 1.00 & -1e-06 & -1e-06 & 3e-06 \\
1986-87 & 0.97 & -5e-05 & -5e-05 & -4e-05 & 0.99 & -4e-05 & -4e-05 & -2e-05 & 0.99 & -4e-05 & -4e-05 & -2e-05 \\
1987-88 & 0.99 & -1e-05 & -1e-05 & -1e-06 & 0.99 & -1e-05 & -1e-05 & -6e-06 & 0.99 & -1e-05 & -1e-05 & -8e-06 \\
1988-89 & 1.00 & -7e-06 & -8e-06 & -4e-06 & 0.99 & -2e-05 & -2e-05 & -1e-05 & 1.00 & -2e-05 & -2e-05 & -1e-05 \\
1989-90 & 1.00 & -1e-05 & -1e-05 & -7e-06 & 0.98 & -2e-05 & -2e-05 & -3e-06 & 0.98 & -4e-05 & -4e-05 & -2e-05 \\
1990-91 & 0.99 & -1e-05 & -1e-05 & -8e-06 & 1.00 & -2e-06 & -4e-06 & -7e-07 & 1.00 & -6e-06 & -8e-06 & -5e-06 \\
1991-92 & 1.00 & 2e-07 & -1e-06 & 5e-06 & 1.00 & -1e-05 & -1e-05 & -1e-05 & 1.00 & -7e-06 & -8e-06 & -4e-06 \\
1992-93 & 1.00 & 4e-07 & -8e-07 & 3e-06 & 0.99 & -8e-06 & -9e-06 & -5e-06 & 1.00 & -8e-06 & -9e-06 & -6e-06 \\
1993-94 & 0.98 & -6e-05 & -6e-05 & -2e-05 & 1.00 & -8e-07 & -2e-06 & 4e-06 & 1.00 & -3e-06 & -5e-06 & 2e-06 \\
1994-95 & 1.00 & -1e-05 & -1e-05 & -5e-06 & 0.99 & -2e-05 & -2e-05 & -9e-06 & 1.00 & -2e-05 & -2e-05 & -6e-06 \\
1995-96 & 1.00 & -1e-05 & -1e-05 & -9e-06 & 1.00 & -7e-06 & -8e-06 & -5e-06 & 1.00 & -8e-06 & -9e-06 & -6e-06 \\
1996-97 & 1.00 & -1e-05 & -2e-05 & -1e-05 & 0.99 & -2e-05 & -2e-05 & -1e-05 & 0.99 & -2e-05 & -2e-05 & -1e-05 \\
1997-98 & 0.98 & -8e-05 & -8e-05 & -2e-05 & 1.01 & -9e-06 & -1e-05 & 6e-06 & 1.00 & -1e-05 & -2e-05 & -8e-06 \\
1998-99 & 0.99 & -1e-05 & -1e-05 & -8e-06 & 0.98 & -4e-05 & -4e-05 & -8e-06 & 1.00 & -1e-05 & -1e-05 & 2e-06 \\
1999-00 & 0.97 & -4e-05 & -4e-05 & -2e-05 & 0.99 & -2e-05 & -2e-05 & -1e-05 & 1.00 & -1e-05 & -1e-05 & -7e-06 \\
2000-01 & 1.00 & -7e-06 & -7e-06 & -6e-06 & 1.00 & -2e-05 & -2e-05 & -1e-05 & 1.00 & -2e-05 & -2e-05 & -1e-05 \\
2001-02 & 1.00 & -4e-06 & -4e-06 & -2e-06 & 1.00 & -4e-06 & -4e-06 & -9e-07 & 1.00 & -3e-06 & -4e-06 & -2e-07 \\
2002-03 & 0.98 & -2e-05 & -2e-05 & -2e-05 & 1.00 & 4e-07 & -5e-07 & 1e-06 & 1.00 & 1e-06 & 4e-08 & 2e-06 \\
2003-04 & 0.97 & -3e-04 & -4e-04 & -7e-06 & 1.00 & -7e-06 & -7e-06 & -5e-06 & 1.00 & -7e-06 & -8e-06 & -5e-06 \\
2004-05 & 1.00 & 1e-06 & 4e-07 & 1e-06 & 1.00 & 4e-06 & 3e-06 & 4e-06 & 1.00 & 4e-06 & 4e-06 & 6e-06 \\
2005-06 & 1.00 & -8e-07 & -1e-06 & -8e-07 & 1.00 & -2e-06 & -2e-06 & -6e-07 & 0.99 & -2e-05 & -2e-05 & -4e-06 \\
2006-07 & 1.00 & 2e-06 & 2e-06 & 3e-06 & 1.00 & -5e-06 & -5e-06 & -3e-06 & 0.99 & -6e-06 & -6e-06 & -4e-06 \\
2007-08 & 1.00 & -5e-06 & -5e-06 & -4e-06 & 1.00 & -1e-06 & -2e-06 & -4e-07 & 1.00 & -8e-08 & -7e-07 & 1e-06 \\
2008-09 & 0.99 & -7e-06 & -8e-06 & -7e-06 & 0.99 & -6e-06 & -7e-06 & -5e-06 & 0.99 & -6e-06 & -7e-06 & -5e-06 \\
2009-10 & 1.00 & 2e-06 & 2e-06 & 2e-06 & 1.00 & -2e-07 & -5e-07 & 1e-06 & 1.00 & -2e-06 & -3e-06 & -1e-06 \\
2010-11 & 1.00 & 8e-07 & 4e-07 & 1e-06 & 1.00 & -9e-07 & -2e-06 & 3e-08 & 1.00 & 3e-07 & -4e-07 & 1e-06 \\
2011-12 & 1.00 & -5e-06 & -5e-06 & -4e-06 & 1.00 & -4e-06 & -4e-06 & -2e-06 & 1.00 & -2e-06 & -2e-06 & -7e-07 \\
2012-13 & 1.00 & 1e-06 & 9e-07 & 2e-06 & 1.00 & -1e-06 & -1e-06 & -1e-07 & 1.00 & -3e-06 & -3e-06 & -2e-06 \\
2013-14 & 0.99 & -4e-06 & -5e-06 & -4e-06 & 1.00 & -7e-06 & -7e-06 & -5e-06 & 1.00 & -7e-06 & -8e-06 & -6e-06 \\
2014-15 & 1.00 & -6e-06 & -6e-06 & -4e-06 & 1.00 & -6e-06 & -7e-06 & -5e-06 & 1.00 & -5e-06 & -6e-06 & -4e-06 \\
2015-16 & 1.00 & -6e-06 & -6e-06 & -6e-06 & 1.00 & -8e-06 & -8e-06 & -7e-06 & 1.00 & -7e-06 & -7e-06 & -6e-06 \\
2016-17 & 1.02 & 4e-06 & 4e-06 & 8e-06 & 1.00 & -4e-06 & -4e-06 & -3e-06 & 1.00 & -3e-06 & -3e-06 & -2e-06 \\
2017-18 & 0.99 & -3e-07 & -5e-07 & -2e-08 & 1.00 & 2e-07 & -2e-08 & 8e-07 & 1.00 & 3e-07 & -7e-08 & 7e-07 \\
2018-19 & 1.00 & -3e-06 & -3e-06 & -3e-06 & 1.00 & -3e-06 & -3e-06 & -2e-06 & 0.99 & -3e-06 & -4e-06 & -2e-06 \\
2019-20 & 1.00 & 6e-06 & 6e-06 & 6e-06 & 1.00 & 3e-06 & 2e-06 & 3e-06 & NA & NA & NA & NA \\
2020-21 & 1.00 & -1e-06 & -1e-06 & -1e-06 & 1.00 & 1e-06 & 1e-06 & 2e-06 & NA & NA & NA & NA \\
\bottomrule
\end{tabular}
\end{table}

\end{document}